\documentclass[12pt]{article}

\IfFileExists{srcltx.sty}{\usepackage[active]{srcltx}}

\usepackage{amssymb,amsmath,bm,array}%
\usepackage[sort&compress,square,comma]{natbib} %
\usepackage{graphicx}%

\catcode`\@=11
\def\sla@#1#2#3#4#5{{%
  \setbox\z@\hbox{$\m@th#4#5$}%
  \setbox\tw@\hbox{$\m@th#4#1$}%
  \dimen4\wd\ifdim\wd\z@<\wd\tw@\tw@\else\z@\fi
  \dimen@\ht\tw@
  \advance\dimen@-\dp\tw@
  \advance\dimen@-\ht\z@
  \advance\dimen@\dp\z@
  \divide\dimen@\tw@
  \advance\dimen@-#3\ht\tw@
  \advance\dimen@-#3\dp\tw@
  \dimen@ii#2\wd\z@
  \raise-\dimen@\hbox to\dimen4{%
    \hss\kern\dimen@ii\box\tw@\kern-\dimen@ii\hss}%
  \llap{\hbox to\dimen4{\hss\box\z@\hss}}}}

\def\slashed#1{%
  \expandafter\ifx\csname sla@\string#1\endcsname\relax
    {\mathpalette{\sla@/00}{#1}}%
  \else
    \csname sla@\string#1\endcsname
  \fi}

\def\declareslashed#1#2#3#4#5{%
  \expandafter\def\csname sla@\string#5\endcsname{%
    #1{\mathpalette{\sla@{#2}{#3}{#4}}{#5}}}}

\catcode`\@=12

\declareslashed{}{/}{.08}{0}{D}
\declareslashed{}{/}{.1}{0}{A}
\declareslashed{}{/}{0}{-.05}{k}
\declareslashed{}{/}{.1}{0}{\partial}
\declareslashed{}{\not}{-.6}{0}{f}

\newcount\hour \newcount\minute \hour=\time \divide \hour by 60 \minute=\time
\def\beq{\begin{equation}}
\def\eeq{\end{equation}}
\def\bea{\begin{eqnarray}}
\def\eea{\end{eqnarray}}

\usepackage{xspace,paralist} %
\usepackage[a4paper,hscale=0.72,vscale=0.72,nohead]{geometry} %

\usepackage[colorlinks=true,bookmarks=false]{hyperref}

\IfFileExists{hypernat.sty}{\usepackage{hypernat}}

\let\oldbox=\Box %
\renewcommand{\Box}{\mathop{\oldbox}}

\let\oldint=\int %
\renewcommand{\int}{\mathop{\hskip -.25ex\oldint \hskip -.4ex}}

\newcommand\gsim{\mathrel{\rlap{\lower4pt\hbox{\hskip1pt$\sim$}}
    \raise1pt\hbox{$>$}}}

\newcommand{\cs}{\textsc{cs}} % Chern-Simons

\newcommand{\p}[1]{\partial_{#1}} %
\newcommand{\la}{\langle} %
\newcommand{\ra}{\rangle} %
\newcommand{\Dsl}{\mathop{\lefteqn{D}{\,\mbox{\large /}}}\nolimits} % Slashed D
\newcommand{\tpsi}{{\tilde\psi}} \newcommand{\tchi}{{\tilde\chi}}
\newcommand{\gev}{~\text{GeV}} %
\newcommand{\tev}{~\text{TeV}} %

\newcommand{\g}[1]{\ensuremath{\gamma^{#1}}}
\newcommand{\uy}{\ensuremath{U_Y(1)}\xspace} %
\newcommand{\ux}{\ensuremath{U_X(1)}\xspace} %

\newcommand{\bc}[1]{c_{#1}} %

\newcommand{\sw}{\sin\theta_w} %
\newcommand{\cw}{\cos\theta_w} %
\newcommand{\ssw}{\sin^2\theta_w} %
\newcommand{\ccw}{\cos^2\theta_w} %

\newcommand{\CA}{\mathcal{A}} %
\newcommand{\CL}{\mathcal{L}} %

\begin{document}

\title{
    \begin{flushright}
      {\small CERN-PH-TH/2009-006}
    \end{flushright}
Anomaly driven signatures of new invisible physics at the Large
    Hadron Collider}

  \author{Ignatios Antoniadis$^{a,b}$, Alexey Boyarsky$^{c,d}$, Sam Espahbodi$^{a,e}$, \\
    Oleg Ruchayskiy$^f$, James D. Wells$^{a,e}$} \date{}
  \maketitle

\begin{it}\noindent
  $^{(a)}$ CERN PH-TH, CH-1211 Geneva 23, Switzerland \\
  $^{(b)}$ CPHT, UMR du CNRS 7644, \'Ecole Polytechnique, 91128 Palaiseau
  Cedex, France \\
  $^{(c)}$ ETHZ, Z\"urich, CH-8093,  Switzerland\\
  $^{(d)}$ Bogolyubov Institute for Theoretical Physics, Kiev 03680, Ukraine\\
  $^{(e)}$ MCTP, University of Michigan, Ann Arbor, MI 48109, USA\\
  $^{(f)}$ \'Ecole Polytechnique F\'ed\'erale de Lausanne, FSB/ITP/LPPC, BSP,
  CH-1015,\\ Lausanne, Switzerland
\end{it}

%\date{\footnotesize\textbf{File \jobname.tex. ${}$Revision: 1.5 ${}$}\\
%  \today. \now}%

%\maketitle

\begin{abstract}
  Many extensions of the Standard Model (SM) predict new neutral vector bosons
  at energies accessible by the Large Hadron Collider (LHC). We study an
  extension of the SM with new chiral fermions subject to non-trivial anomaly
  cancellations.  If the new fermions have SM charges, but are too heavy to be
  created at LHC, and the SM fermions are not charged under the extra gauge
  field, one would expect that this new sector remains completely invisible at
  LHC. We show, however, that a non-trivial anomaly cancellation between the
  new heavy fermions may give rise to observable effects in the gauge boson
  sector that can be seen at the LHC and distinguished from backgrounds.

\end{abstract}

\vfill\eject

\tableofcontents

%%%%%%%%%%%%%%%%%%%%%%%%%%%%%%%%%%%%
\section{Introduction: Mixed Anomalies in Gauge Theory}
\label{sec:introduction-1}

It is well known that theories in which fermions have chiral couplings with
gauge fields suffer from \emph{anomalies} -- a phenomenon of breaking of gauge
symmetries of the classical theory at one-loop level. Anomalies make a theory
inconsistent (in particular, its unitarity is lost). The only way to restore
consistency of such a theory is to arrange the exact \emph{cancellation} of
anomalies between various chiral sectors of the theory.  This happens, for
example, in the Standard Model (SM), where the cancellation occurs between quarks
and leptons within each
generation~\cite{Gross:1972pv,Bouchiat:1972iq,Georgi:1972bb}.

Another well-studied example is the Green-Schwarz anomaly cancellation
mechanism~\cite{Green:84} in string theory. In this case the cancellation
happens between the anomalous contribution of chiral matter of the closed
string sector with that of the open string.\footnote{Formally, the
  Green-Schwarz anomaly cancellation occurs due to the anomalous Bianchi
  identity for the field strength of a 2-form closed string. However, this
  modification of Bianchi identity arises from the 1-loop contribution of
  chiral fermions in the open string sector. A toy model, describing
  microscopically Green-Schwarz mechanism was studied e.g.
  in~\cite{Boyarsky:02}.}

Particles involved in anomaly cancellation may have very different
masses -- for example, the mass of the top quark in the SM is much higher than
the masses of all other fermions.  On the other hand, gauge invariance should
pertain in the theory at all energies, including those which are smaller than the
mass of one or several particles involved in anomaly cancellation.  The usual
logic of renormalizable theories tells us that the interactions, mediated by
heavy fermions running in loops, are generally suppressed by the masses of
these fermions~\cite{Appelquist:74}. The case of anomaly cancellation presents
a notable counterexample to this famous ``decoupling theorem'' -- the
contribution of \emph{a priori} arbitrary heavy particles should remain
unsuppressed at arbitrarily low energies.  As was pointed out by D'Hoker
and Farhi~\cite{DHoker:84a,DHoker:84b}, this is possible because anomalous
(i.e. gauge-variant) terms in the effective action have topological nature and are
therefore scale independent.  As a result, they are not suppressed
even at energies much smaller than the masses of the particles producing
these terms via loop effects.
This gives hope to see some signatures at low energies 
generated by new high-energy physics.

One possibility is to realize non-trivial anomaly cancellation in the
electroweak (EW) sector of the SM. Here the electromagnetic $U(1)$ subgroup is not
anomalous by definition.  However, the mixed triangular hypercharge
%\uy group, mixed 
$\uy\times SU(2)^2$ anomalies and gravitational anomalies are non-zero for a
generic choice of hypercharges. If one takes the most general choice of
hypercharges, consistent with the structure of the Yukawa terms, one sees that
it is parametrized by two independent quantum numbers $Q_e$ (shift of
hypercharge of left-handed lepton doublet from its SM value) and $Q_q$
(corresponding shift of quark doublet hypercharge). All the anomalies are then
proportional to \emph{one} particular linear combination: %of these numbers:
$\epsilon = Q_e+3Q_q$.  Interestingly enough, $\epsilon$ is equal to the sum
of \emph{electric charges} of the electron and proton.  The experimental upper
bound on the parameter $\epsilon$, coming from checks of electro-neutrality of
matter is rather small: $\epsilon < 10^{-21}e$~\cite{Marinelli:83,PDG:06}.  If
it is non-zero, the anomaly of the SM has to be cancelled by additional
anomalous contributions from some physics beyond the SM, possibly giving rise
to some non-trivial effects in the low energy effective theory.

In the scenario described above the anomaly-induced effects are
proportional to a very small parameter, which makes experimental
detection very difficult. In this paper we consider another situation, where
anomalous charges and therefore, anomaly-induced effects, are of order one.
To reconcile this with existing experimental bounds, such an anomaly
cancellation should take place between the SM and ``hidden'' sector, with the
corresponding new particles appearing at relatively high energies.  Namely,
many extensions of the SM add extra gauge fields to the SM gauge group
(see e.g.~\cite{Hewett:88} and refs.\ therein).  For example, additional $U(1)$s
naturally appear in models in which $SU(2)$ and $SU(3)$ gauge factors of
the SM arise as parts of unitary $U(2)$ and $U(3)$ groups (as e.g.\ in D-brane
constructions of the SM~\cite{Ibanez:2001nd,Antoniadis:02,Anastasopoulos:06}).
In this paper, we consider extensions of the SM with an
additional \ux factor, so that the gauge group becomes $SU(3)_c \times SU(2)_W
\times \uy\times \ux$.  As the SM fermions are chiral with respect to the EW
group $SU(2)_W \times \uy$, even choosing the charges for the \ux group
so that the triangular $\ux^3$ anomaly vanishes,  mixed anomalies may still arise: 
$\ux\uy^2$, $\ux^2\uy$, $\ux SU(2)^2$. In this
work we are interested in the situation when only (some of these) \emph{mixed
  anomalies} with the electroweak group $SU(2)\times \uy$ are non-zero.  A
number of works have already discussed such theories and their signatures (see
e.g.~\cite{Antoniadis:06,Anastasopoulos:06,Coriano:07a,Armillis:07a,Coriano:07b,Kumar:07,Antoniadis:07a}).

The question of experimental signatures of such theories at the LHC should be
addressed differently, depending on whether or not the SM fermions are charged
with respect to the \ux group:

%\begin{figure}[t]
 % \centering
 % \includegraphics[width=.5\linewidth,height=.2\textheight]{diagram-qq-X-ff}%
 % \caption{Direction production of neutral $X$ boson in $pp$ collision.}
  %\label{fig:qq-X-ff}
%\end{figure}

\begin{asparaitem}
\item If SM fermions are charged with respect to the \ux group, and the mass
  of the new $X$ boson is around the TeV scale, we should be able to see the
  corresponding resonance in the forthcoming runs of LHC in e.g., $q\bar q\to
  X\to f\bar f$.  The analysis of this is rather standard $Z'$ phenomenology,
  although in this case an important question is to distinguish between
  theories with non-trivial cancellation of mixed anomalies, and those that
  are anomaly free.
% \begin{compactenum}[\bf 1.]
% \item\label{item:1} SM fermions are {not} charged w.r.t. \ux
% \item SM fermions are vector-like w.r.t. \ux (thus SM Higgs is \ux neutral)
% \item SM fermions are chiral w.r.t. \ux (and therefore SM Higgs is \ux charged
%   and the pattern of EW symmetry breaking gets modified)
% \end{compactenum}
\item On the other hand, one is presented with a completely different
  challenge if the SM fermions are \emph{not charged} with respect to the \ux
  group. This makes impossible the usual direct production of the $X$ boson
  via coupling to fermions. Therefore, the question of whether an anomalous
  gauge boson with mass $M_X \sim 1$~TeV can be detected at LHC becomes
  especially interesting.
\end{asparaitem}

A theory in which the cancellation of the mixed $\ux SU(2)^2$ anomaly occurs
between some heavy fermions and Green-Schwarz (i.e. \emph{tree-level
  gauge-variant}) terms was considered in~\cite{Kumar:07}.  The leading
non-gauge invariant contributions from the triangular diagrams of heavy
fermions, unsuppressed by the fermion masses, cancels the Green-Schwarz
terms.  The triangular diagrams also produce subleading (gauge-invariant)
terms, suppressed by the mass of the fermions running in the loop. This leads to an
appearance of dimension-6 operators in the effective action, having the
general form ${F_{\mu\nu}^3}/{\Lambda_X^2}$, where $F_{\mu\nu}$ is the
field strength of $X$, $Z$ or $W^\pm$ bosons. Such terms contribute to the
$XZZ$ and $XWW$ vertices. As the fermions in the loops are heavy, such
vertices are in general strongly suppressed by their mass.  However, motivated
by various string constructions, \cite{Kumar:07} assumed two things:
\emph{(a)} these additional massive fermions are above the LHC reach but not
too heavy (e.g.,  have masses in tens of TeV); \emph{(b)} there are many such
fermions (for instance Hagedorn tower of states) and therefore the mass suppression 
%of fermions $\Psi$ 
can be compensated by the large multiplicity of these fermions.

In this paper we consider another possible setup, in which the anomaly
cancellation occurs \emph{only within a high-energy sector} (at scales
not accessible by current experiments), but at low energies there remain
contributions unsuppressed by masses of heavy particles. A similar setup, with
completely different phenomenology, has been previously considered
in~\cite{Antoniadis:06,Antoniadis:07a}.

The paper is organized as follows. We first consider in
section~\ref{sec:gauge-invar-chern} the general theory of D'Hoker-Farhi terms
arising from the existence of heavy states that contribute to anomalies. We
illustrate the theory issues in the following section~\ref{sec:toy-example}
with a toy model.  In section~\ref{sec:choic-charg-real} we give a complete
set of charges for a realistic $SU(2)\times U(1)_Y\times U(1)_X$ theory. In
section~\ref{sec:phenomenology} we bring all these elements together to
demonstrate expected LHC phenomenology of this theory.

%\section{\ux neutral SM fermions}
%\label{sec:work-wells-et}

%%%%%%%%%%%%%%%%%%%%%%%%%%%%%%%%%%%%%%%%%%%
\section{D'Hoker-Farhi Terms from High Energies}
\label{sec:gauge-invar-chern}

In this Section we consider an extension of the SM with an additional \ux
field. The SM fields are neutral with respect to the \ux group, however, the
heavy fields are charged with respect to the electroweak (EW) $\uy\times
SU(2)$ group. This leads to a non-trivial mixed anomaly cancellation in the
heavy sector and in this respect our setup is similar to the
work~\cite{Kumar:07}.  However, unlike the work~\cite{Kumar:07}, we show that
there exists a setup in which non-trivial anomaly cancellation induces a
\emph{dimension 4 operator} at low energies.  The theories of this type were
previously considered in~\cite{Antoniadis:06,Antoniadis:07a}.

At energies accessible at LHC and below the masses of the new heavy
fermions, the theory in question is simply the SM plus a massive vector boson
$X$:
\begin{equation}
\label{eq:39}
\mathcal{L} = \mathcal{L}_{SM} - \frac1{4g_X^2} |F_X|^2 + \frac{M_X^2}2
|D\theta_X|^2  + \CL_{int}
\end{equation}
where $\theta_X$ a pseudo-scalar field, charged
under \ux so that $D\theta_X = d\theta_X + X$ remains gauge invariant
(St\"uckelberg field). One can think about $\theta_X$ as being a phase of
a heavy Higgs field, which gets ``eaten'' by the longitudinal component of the
$X$ boson. Alternatively, $\theta_X$ can be a component of an
antisymmetric $n$-form, living in the bulk and wrapped around an $n$-cycle.
The interaction term $\CL_{int}$ contains the vertices between the $X$ boson
and the $Z,\gamma,W^\pm$:
\begin{equation}
  \label{eq:59}
  \epsilon^{\mu\nu\lambda\rho} Z_\mu X_\nu \p\lambda Z_\rho,\qquad   \epsilon^{\mu\nu\lambda\rho} Z_\mu X_\nu \p\lambda \gamma_\rho,\qquad   \epsilon^{\mu\nu\lambda\rho} W^+_\mu X_\nu \p\lambda W^-_\rho, 
\end{equation}
We wish to generalize these terms into an $SU(2)\times \uy$ covariant form. One
possible way would be to have them arise from
\begin{equation}
  \label{eq:64}
  \epsilon^{\mu\nu\lambda\rho} X_\mu Y_\nu  \p\lambda
  Y_\rho\quad\text{and}\quad \epsilon^{\mu\nu\lambda\rho} X_\mu \omega_{\nu\lambda\rho}(A^a)
\end{equation}
where $\omega_{\nu\lambda\rho}(A^a)$ is the Chern-Simons term, built of the  $SU(2)$ fields
$A_\mu^a$:
\begin{equation}
  \label{eq:90}
  \omega_{\nu\lambda\rho}(A^a) = A^a_\nu \p\lambda A_\rho^a + \frac23 \epsilon_{abc} A^a_\nu A_\lambda^b A_\rho^c
\end{equation}
%Expressions~(\ref{eq:64}) are gauge-invariant. 
However, apart from the
desired terms of eq.~(\ref{eq:59}) they contain also terms like $
\epsilon^{\mu\nu\lambda\rho} X_\mu \gamma_\nu \p\lambda \gamma_\rho $ which is
not gauge invariant with respect to the electro-magnetic $U(1)$ group, and thus unacceptable.

To write the expressions of~(\ref{eq:59}) in a gauge-covariant form, we should
recall that it is the SM Higgs field $H$ which selects massive directions
through its covariant derivative $D_\mu H$. Therefore, we can write the
interaction term in the following, explicitly $SU(2)\times\uy\times \ux$
invariant, form:
\begin{equation}
  \label{eq:48}
  \CL_{int} = \bc1 \frac{H^\dagger D H}{|H|^2} D\theta_X F_Y+\bc2 \frac{H F_W D H^\dagger}{|H|^2} D\theta_X 
\end{equation}
%where $H$ is the SM Higgs field, 
The coefficients $\bc1,\bc2$ are dimensionless and can have arbitrary values,
determined entirely by the properties of the high-energy theory.
In eq.~(\ref{eq:48}) we use the differential form notation (and further we omit the
wedge product symbol $\wedge$) to keep expressions more compact. We will often
call the terms in eq.~(\ref{eq:48}) as the \emph{D'Hoker-Farhi}
terms~\cite{DHoker:84a,DHoker:84b}.

What can be the origin of the interaction terms~(\ref{eq:48})? The simplest
possibility would be to add to the SM several heavy fermions, charged with
respect to $SU(2)\times \uy\times \ux$. Then, at energies below their masses
the terms (\ref{eq:48}) will be generated.
Below, we illustrate this idea in a toy-model setup.

 \begin{table}[t]
  \centering
  \begin{tabular}[c]{%|>{$}c<{$}|>{$}c<{$}|
                     |>{$}c<{$}||>{$}c<{$}|>{$}c<{$}|>{$}c<{$}|>{$}c<{$}
                    ||>{$}c<{$}|>{$}c<{$}|>{$}c<{$}|>{$}c<{$}|>{$}c<{$}|} 
    \hline
    & \multicolumn{2}{c|}{$\psi_1$} & \multicolumn{2}{c||}{$\psi_2$} &
    \multicolumn{2}{c|}{$\chi_1$} & \multicolumn{2}{c|}{$\chi_2$} \\
    \hline
%    & \phi_1 & \phi_2 & 
    &\psi_{1L} & \psi_{1R}& \psi_{2L} & \psi_{2R} &\chi_{1L} &
    \chi_{1R} & \chi_{2L} & \chi_{2R}\\
    \hline
    U(1)_A 
%& \k{} & 0 
    & e_1 & e_1 & e_2 & e_2 & e_4 & e_3 & e_3 &
    e_4\\
    \hline
    U(1)_B & 
%0 & -\k{} &
    q_{1} & -q_{1}  &  -q_1 & q_1 & q_2 & q_2 & -q_2 & -q_2
    \\\hline
  \end{tabular}
  \caption{A simple choice of charges for all fermions, leading
    to the low-energy effective action~(\ref{eq:97}). The charges are
    chosen in such a way that all gauge anomalies cancel. The cancellation of
    $U(1)_A^3$ and $U(1)_B^3$ anomalies happens for any value of
    $e_i,q_i$. Cancellation of mixed anomalies requires 
    $q_2=\displaystyle \frac{q_1(e_1^2-e_2^2)}{2(e_3^2-e_4^2)}$.} 
  \label{tab:toymodel}
\end{table}

Consider a theory with a set of chiral fermions $\psi_{1,2}$ and
$\chi_{1,2}$, charged with respect to the gauge groups $U(1)_A\times U(1)_B$.
As the fermions are chiral, they can obtain masses only through Yukawa
interactions with both $\Phi_1$ and $\Phi_2$ scalar fields. $\Phi_{1}$ is charged with respect to
$U(1)_B$, and $\Phi_2$ is charged with respect to $U(1)_A$:
\begin{align}
  \CL_{Yukawa} &= i\sum_{i=1,2}\bar \psi_i \Dsl \psi_i + (f_1v_1) \bar\psi_1
  e^{i\g5 \theta_B} \psi_1 + (f_2v_2) \bar\psi_2
  e^{-i\g5 \theta_B }\psi_2 \notag\\
  &+ i\sum_{i=1,2}i\bar \chi_i \Dsl \chi_i + (\lambda_1 v_2) \bar\chi_1
  e^{i\gamma_5 \theta_A} \chi_1 + (\lambda_2 v_2) \bar\chi_2 e^{-i\gamma_5
    \theta_A} \chi_2 +\mathrm{h.c.}  \label{eq:12}
\end{align}
Here we have taken $\Phi_1$ in the form $\Phi_1 = v_1 e^{i\theta_B}$, where
$v_1$ is its vacuum expectation value (VEV) and $\theta_B$ is charged with respect
to the $U(1)_B$ group with charge $2q_1$, and $\Phi_2 = v_2 e^{i\theta_A}$,
where $\theta_A$ is charged with respect to $U(1)_A$ group with charge
$e_3 - e_4$. 

\begin{figure}[t]
  \centering
  \includegraphics{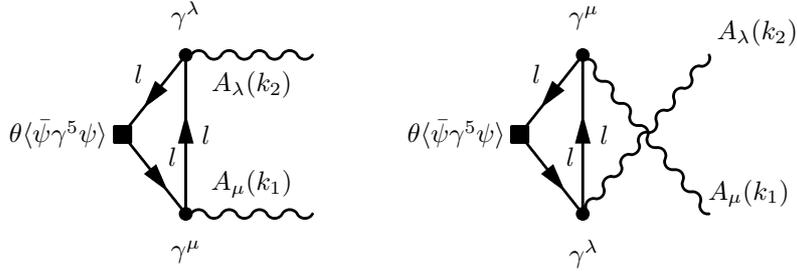} %
  \caption{Anomalous contributions to the correlator $\la\bar \psi \gamma^5 \psi\ra$%,
%    denoted by $I_\text{\protect\ref{fig:diagram-triangular-theta}a}$ and
%$I_\text{\protect\ref{fig:diagram-triangular-theta}b}$ in the text
    .} %
  \label{fig:diagram-triangular-theta}
\end{figure}

\begin{figure}[t]
  \centering %
%  \begin{minipage}[c]{.45\textwidth}
% \includegraphics{diagram-triangular-1} %
%  \end{minipage}~\begin{minipage}[c]{.45\textwidth}
% \includegraphics{diagram-triangular-2} %
  \includegraphics{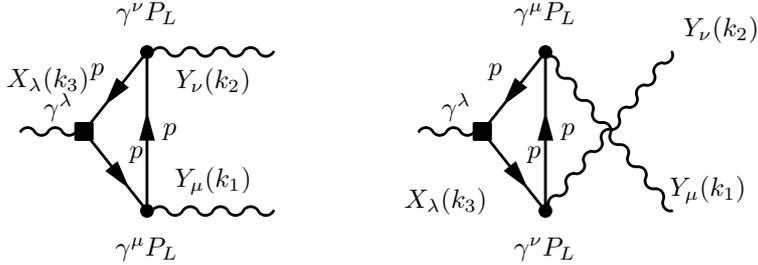}
%  \end{minipage}
  \caption{Two graphs, contributing to the Chern-Simons terms} %
  \label{fig:diagram-triangular} %
\end{figure}

The structure of the Yukawa terms restricts the possible charge assignments,
so that the fermions $\psi_{1,2}$ should be vector-like with respect to the
group $U(1)_A$ and chiral with respect to the $U(1)_B$ (and vice versa for the
fermions $\chi_{1,2}$).  The choice of the charges in Table~\ref{tab:toymodel}
is such that triangular anomalies $[U(1)_A]^3$ and $[U(1)_B]^3$ cancel
separately for the $\psi$ and $\chi$ sector for any choice of $e_i,q_i$. The
cancellation of mixed anomalies occurs only \emph{between} $\psi$ and $\chi$
sectors. It is instructive to analyze it at energies below the masses of all
fermions. The terms in the low-energy effective action, not suppressed by the
scale of fermion masses are given by
\begin{equation}
  \label{eq:1008}
  S_\cs =\int \frac{(e_1^2 -e_2^2)q_1}{16\pi^2} \theta_B F_A\wedge F_A +
  \frac{(e_3^2 -e_4^2)(2q_2)}{16\pi^2} \theta_A F_A\wedge F_B +
  \alpha
  A\wedge  B\wedge F_A 
\end{equation}
The diagrammatic expressions for the first two terms are shown in
Fig.~\ref{fig:diagram-triangular-theta}, while the Chern-Simons (CS) term is
produced by the diagrams of the type presented in
Fig.~\ref{fig:diagram-triangular}. The contribution to the CS term $A\wedge
B\wedge F_A$ comes from both sets of fermions.  Only fermions $\psi$
contribute to the $\theta_B$ terms and only fermions $\chi$ couple to
$\theta_A$ and thus contribute to $\theta_A F_A\wedge F_B$.  Notice that
while coefficients in front of the $ \theta_A$ and $\theta_B$ terms are
uniquely determined by charges, the coefficient $\alpha$ in front of the CS
term is \emph{regularization dependent}. As the theory is anomaly free, there
exists a choice of $\alpha$ such that the expression~(\ref{eq:1008}) becomes
gauge-invariant with respect to both gauge groups. Notice, however, that in
the present case $\alpha$ cannot be zero, as $\theta_A F_A\wedge F_B$ and $
\theta_B F_A\wedge F_A$ have gauge variations with respect to different
groups.  For the choice of charges presented in Table~\ref{tab:toymodel},
the choice of 
$\alpha$ is restricted such that expression~(\ref{eq:1008}) can be written in an
explicitly gauge-invariant form:
\begin{equation}
  \label{eq:97}
    S_\cs =\int \kappa D\theta_A \wedge D\theta_B\wedge F_A 
\end{equation}
where the relation between the coefficient $\kappa$ in front of the CS term
and the fermion charges is given by
\begin{equation}
\label{eq:364}
\alpha\equiv \kappa = \frac{q_1 (e_1^2 - e_2^2)}{16\pi^2}
\end{equation}
For the anomaly cancellation, it is also necessary to impose the condition 
\begin{equation}
q_2=\frac{q_1(e_1^2-e_2^2)}{2(e^2_3-e^2_4)}
\end{equation}
as indicated in table~\ref{tab:toymodel}.

The term~(\ref{eq:97}) was obtained by integrating out heavy fermions
(Table~\ref{tab:toymodel}). The resulting expression is not suppressed by
their mass and contains only a dimensionless coupling $\kappa$. Unlike the case
of~\cite{DHoker:84a,DHoker:84b}, the anomaly was cancelled entirely among the
fermions which we had integrated out.  The expression~(\ref{eq:97}) represents
therefore an apparent counterexample of the ``decoupling
theorem''~\cite{Appelquist:74}. Note that the CS term~(\ref{eq:97})
contains only \emph{massive} vector fields. This effective action can only be
valid at energies \emph{above} the masses of all vector fields and
\emph{below} the masses of all heavy fermions, contributing to it.  However,
masses of both types arise from the same Higgs fields.  Therefore a hierarchy
of mass scales can only be achieved by making gauge couplings smaller
than Yukawa couplings.  On the other hand, the CS coefficient $\kappa$ is
proportional to the (cube of the) gauge couplings. Therefore we can
schematically write a dimensionless coefficient $\kappa \sim
({M_V}/{M_f})^3$, where $M_V$ is the mass of the vector fields and $M_f$ is
the mass of the fermions (with their Yukawa couplings $\sim 1$). In the
limit when $M_f$ is sent to infinity, while keeping $M_V$ finite, the
decoupling theorem holds, as the CS terms get suppressed by the small gauge
coupling constant.  However, a window of energies $M_V \lesssim E \lesssim
M_f$, at which the term~(\ref{eq:97}) is applicable, always remains and this
opens interesting phenomenological possibilities, which are absent in the
situation when the corresponding terms in the effective action are suppressed as $E/M_f$
(as in~\cite{Appelquist:74}) and not as $M_V/M_f$.

Finally, it is also possible that the fermion masses are not generated via the
Higgs mechanism, (e.g. coming from extra dimensions) and are not directly
related to the masses of the gauge fields. In this case, the decoupling theorem
may not hold and new terms can appear in a wide range of energies (see
e.g.~\cite{anomaly-th,anomaly-exp} for discussion).

\noindent 
% The Higgs field $\Phi$ which provides mass to the fermions $\Psi_i$ is charged
% with respect to the $U(1)_B$ with charge $2q_1$. If it acquires a vacuum
% expectation value (VEV) $|\Phi_1| =\vb$, one has $\Phi_1 = \vb e^{2iq\theta}$
% and the Yukawa term for $\Psi$ fields becomes $\vb\bar\Psi_i
% e^{2iq_1\theta\g5} \Psi_i$. The Higgs then provides mass to the $B_\mu$-field:
% \begin{equation}
%   \label{eq:375}
%   \mb = 2 q_1 \vb
% \end{equation}
% and Higgs's kinetic term becomes that of the field $\theta$ in
% action~(\ref{eq:4}).

%%%%%%%%%%%%%%%%%%%%%%%%%%%%%%%%%%%%%%%%
\section{A Standard Model Toy Example}
\label{sec:toy-example}

Let us now generalize this construction to the case of interest, when one of
the scalar fields generates mass for the chiral fermions and is the SM Higgs
field, while at the same time the masses of all new fermions are higher than about $10$~TeV.

Note that previously, in the theory described by~(\ref{eq:12}) the mass terms for fermions
were diagonal in the basis $\psi,\chi$ and schematically had the form $m_1
\bar\psi\psi + m_2 \bar\chi\chi$. To make both masses for $\psi$ and $\chi$
heavy (i.e., determined by the non-SM scalar field), while still preserving a
coupling of the fermions with the SM Higgs, we consider a non-diagonal mass term which
(schematically) has the following form:
\begin{equation}
  \label{eq:62}
  \CL_{mass} = m\bar\psi \psi + M(\bar \psi \chi +  \bar\chi \psi)
\end{equation}
Computing the eigenvalues of the mass matrix, we find that the two mass
eigenstates have masses $M\pm \frac m2$ (in the limit $m\ll M$).  
%The change
%to the basis $\tilde \psi$, in which the mass terms are diagonal, is given by
%\begin{equation}
%  \label{eq:63}
 % \left\{%
   % \begin{aligned}
     % \psi & = \hphantom{-} \cos\theta \tilde\psi_1 + \sin\theta (\gamma_5\tilde\psi_2)\\
      %\chi & = -\sin\theta \tilde\psi_1 + \cos\theta (\gamma_5\tilde\psi_2)\\
    %\end{aligned}\right.
%\end{equation}
%with the angle $\theta$ given by
%\begin{equation}
  %\label{eq:66}
  %\cos\theta = \frac1{\sqrt 2}\left(1+\frac m{2M}\right)\quad;\quad \sin\theta = \frac1{\sqrt 2}\left(1-\frac m{2M}\right)
%\end{equation}
%The mass term becomes:
 %\begin{equation}
   %\label{eq:65}
   %\CL_\text{mass} = M_1\bar{\tilde{\psi_1}} \tilde\psi_1 + M_2\bar{\tilde{\psi_2}} \tilde\psi_2
 %\end{equation}
%where
%\begin{equation}
  %\label{eq:74}
 % M_{1,2} = \sqrt{M^2 + \frac{m^2}4}\pm \frac{m}2\approx M \pm \frac m2
%\end{equation}
%for $m\ll M$.
%%%%#################################
% we find for the values of $\cos\theta$ and $\sin\theta$:

% The interaction of $\psi_1$ with the Higgs field is given by
% \begin{equation}
%   \label{eq:96}
%   m\bar\psi_1 e^{i\gamma_5 \theta_H}\psi_1\approx  m\bar\psi_1 \psi_1 + i m
%   \theta_H \bar\psi_1 \gamma_5 \psi_1
% \end{equation}
% In terms of the mass eigenvalues 
% \begin{equation}
%   \label{eq:98}
%   i m
%   \theta_H \bar\psi_1 \gamma_5 \psi_1 = im\theta_H (\bar\chi_1 \gamma_5\chi_1
%   - \bar\chi_2 \gamma_5\chi_2)
% \end{equation}

% \subsubsection{``Mixed'' mass term for Chiral fermions}
% \label{sec:charged-fermions}

Now, we consider the case when the mass terms, similar to those of
Eq.~(\ref{eq:62}) are generated through the Higgs mechanism.  We introduce two
complex scalar Higgs fields: $H=H_1 + i H_2$ and $\Phi = \Phi_1 + i \Phi_2$.
$H$ is charged with respect to the \uy only (with charge $1$), while
$\Phi$ is \uy neutral, but has charge $1$ with respect to the \ux.
We further assume that both Higgs fields develop non-trivial VEVs:
\begin{equation}
  \label{eq:61}
  \langle H \rangle = v\quad ; \quad \langle \Phi\rangle = V\quad ; \quad v \ll V
\end{equation}
Then, we may write
\begin{equation}
  \label{eq:69}
  H = v e^{i\theta_H}; \qquad \Phi = V e^{i\theta_X}
\end{equation}
neglecting physical Higgs field excitations ($H(x) = (v+h(x))e^{i\theta_H}$,
etc.).

Let us suppose that the full gauge group of our theory is just $\uy\times
\ux$. Consider 4 Dirac fermions ($\psi_1,\psi_2,\chi_1,\chi_2$) with the following
% (see
% Table~\ref{tab:2ferm})\footnote{We work in the basis of $\gamma$-matrices,
%   where $\g5=\mathrm{diag}(1,-1)$ and left (right) projectors are given by
%   $P_{L,R} = \frac12(1\pm \g5)$.}
\begin{table}
  \centering
  \begin{tabular}{>{$}c<{$}|>{$}c<{$}|>{$}c<{$}|>{$}c<{$}|>{$}c<{$}|>{$}c<{$}|>{$}c<{$}|>{$}c<{$}|>{$}c<{$}}
    \hline
    ~ & \multicolumn{2}{c|}{$\psi_1$} & \multicolumn{2}{c|}{$\psi_2$} &
    \multicolumn{2}{c|}{$\chi_1$} & \multicolumn{2}{c}{$\chi_2$}\\
    \hline
    & \psi_{1L} & \psi_{1R} & \psi_{2L} & \psi_{2R} & \chi_{1L} & \chi_{1R} & \chi_{2L} & \chi_{2R}\\
    \hline
    Q_X & x & x & x & x& x-1& x+1& x+1 & x-1\\
    \hline
    Q_Y & y & y+1 & y_1 & y_2 & y+1 & y & y_2 & y_1\\
    % \cline{2-9}
    % \epsilon = -1\;;\; y_1 = y+1&  y & y+1 & y+1  & y & y+1 & y & y & y+1\\
    \hline
  \end{tabular}
  \caption{Charge assignment for the $\uy\times\ux$ with 4 Dirac
    fermions. Charges of the scalar fields $H$ and $\Phi$ are equal to (1,0) and (0,1), respectively.}
  \label{tab:2ferm}
\end{table}
% These charge assignment allow to write the
Yukawa terms, leading to the
Lagrangian in the form, similar to~(\ref{eq:62}):
\begin{equation}
  \label{eq:70}
  \CL_\text{Yukawa} = %\sum_{i=1,2} 
  m_1  \bar\psi_1  e^{i\g5\theta_H} \psi_1
  % +  m_2 \bar\psi_2
  % e^{i\epsilon\g5\theta_H} \psi_2
+ M_1 (\bar \psi_1 e^{i\g5\theta_X} \chi_1 + c.c.) + M_2 (\bar \psi_2 e^{-i\g5\theta_X} \chi_2 + c.c.)
\end{equation}
Here we introduced masses $m_{1}=f_{1} v$ and $M_{1,2}=F_{1,2} V$, with
$f_{1}$ and $F_{1,2}$ the corresponding Yukawa couplings.
% , such that
% $m_{1,2}\ll M_{1,2}$ and put $\epsilon=\pm 1$, to account for two possible
% ways to couple $\psi_2$ with $H$. This term immediately implies that $y_2 =
% y_1 + \epsilon$

The choice of fermion charges is dictated by the Yukawa terms~(\ref{eq:70}).
The $\psi$ fermions are vector-like with respect to \ux group, although chiral
with respect to the \uy. The fermions $\psi_1,\chi_1$ (and $\psi_2,\chi_2$)
have charges with respect to \uy group, such that
\begin{equation}
  \label{eq:71}
  Q_Y(\psi_{1L}) = Q_Y(\chi_{1R}) \quad\text{and} \quad Q_Y(\psi_{1R}) =
  Q_Y(\chi_{1L}) 
\end{equation}
and similarly for the pair $\psi_2,\chi_2$. Unlike $\psi_1$, the fermions
$\psi_2$ do not have Yukawa term $m_2 \bar\psi_2 e^{i\epsilon\g5\theta_H}
\psi_2$, as this would make the choice of charges too restrictive and does not
allow us to generate terms similar to~(\ref{eq:97}). The resulting charge
assignment is shown in Table~\ref{tab:2ferm}.

% \subsubsection{Anomalies in theory~(\ref{eq:62})}
% \label{sec:anomalies-theory}

% Let us analyze the anomaly structure of the theory with the choice of
% charges given by the Table~\ref{tab:charges}.

It is clear that the triangular anomalies $XXX$ and $YYY$ cancel as there is
equal number of left and right moving fermions with the same charges. Let us
consider the mixed anomaly $XYY$. The condition for anomaly cancellation is given
by
\begin{align}
  \label{eq:72}
  \mathcal{A}_{XYY} &= \sum Q_X^L (Q_Y^L)^2 - Q_X^R (Q_Y^R)^2=y_1^2 + y_2^2 -
  1 - 2y -2y^2 = 0
% \\
%   & = \left\{
%     \begin{aligned}
%       -2 (1 + y - y_1) (y + y_1) &\quad& \epsilon &= -1\\
%       -2 (1 + y + y_1) (y - y_1) &\quad& \epsilon &= +1
%     \end{aligned}\right.
\end{align}
The other mixed anomaly $XXY$ is proportional to
\begin{equation}
  \label{eq:73}
  \mathcal{A}_{XXY} = 1-{y_1}+{y_2}+2 {x} (-2
  y+{y_1}+{y_2}-1) =0
% \left\{
%     \begin{aligned}
%       -4 x (1 + y - y_1), &\quad& \epsilon &= -1\\
%       2 - 4 x( y - y_1), &\quad& \epsilon &= 1
%     \end{aligned}\right.
\end{equation}
and should also cancel.
% Both anomalies cancel for the following choice of charges:
% \begin{equation}
%   \label{eq:94}
%   \begin{aligned}
%     y_2 & = y_1 + \epsilon & y_1 & = y+1 & \epsilon & = -1\\
%     y_2 & = y_1 + \epsilon & y_1 & = -y-1 \quad;\quad 2x(2y+1)=-1& \epsilon &
%     = 1
%   \end{aligned}
% \end{equation}
% Both choices (for $\epsilon=\pm1$) cancel XYY and XXY anomalies
% \emph{separately} in $\psi_{1,2}$ and $\chi_{1,2}$ sectors. This is of course
% a legitimate choice of fermion charges which leads to an anomaly-free theory.
% By integrating out these fermions, one will be left with the dimension six
% operators, similar to those, considered in~\cite{Kumar:07} (if masses $M_1\neq
% M_2$ and $m_1 \neq m_2$). 

In analogy with the toy-model, described above, Table~\ref{tab:choice}
presents an anomaly free assignment for which the mixed anomalies cancel only
between the $\psi$ and $\chi$ sectors and lead to the following term in the
effective action (similar to~(\ref{eq:97})):
\begin{equation}
  \label{eq:91}
  \CL_\CA = \kappa D\theta_H \wedge D \theta_X \wedge F_Y
\end{equation}
% in the low-energy effective action 
Here the parameter $\kappa$ is defined by the $XYY$ anomaly in the $\psi$ or
$\chi$ sector, in analogy with Eq.~(\ref{eq:364}):
\begin{equation}
  \label{eq:92}
  \kappa =-\frac{x \left(-y_1^2+y_2^2+2 y+1\right)}{32\pi^2}
\end{equation}
To have $\kappa\neq 0$ we had to make two
mass eigenstates in the sector $\psi_2,\chi_2$  degenerate and equal to
$M_2$. The charges $x, y$ become then arbitrary, while $y_{1,2}$ should satisfy the
constraints~(\ref{eq:72}) and (\ref{eq:73}). It is easy to see that indeed this can be done
%constraints~(\ref{eq:72}) and (\ref{eq:73}) can be satisfied, 
together with the inequality $\kappa\neq 0$. The solution gives:
\begin{equation}
  \label{eq:93}
  y_1 =\frac{4 y x^2-4 y x-4 x-y}{4
    x^2+1}\quad;\quad y_2 = \frac{4 y x^2+4 x^2+4 y x-y-1}{4
    x^2+1}
\end{equation}
The choice~(\ref{eq:93}) leads to the following value of $\kappa$:
\begin{equation}
  \label{eq:95}
  \kappa = -\frac{2 x \left(4 x^2-1\right) \left((8
      y+4) x^2+8 y (y+1) x-2
      y-1\right)}{\left(4 x^2+1\right)^2}
\end{equation}
One can easily see that $\kappa$ is non-zero for generic choices of $x$ and
$y$. One such a choice is shown in Table~\ref{tab:choice} (recall that all
\ux charges are normalized so that $\theta_X$ has $Q_X(\theta_X)=1$ and all
\uy charges are normalized so that $Q_Y(\theta_H)=1$).
\begin{table}
  \centering
  \begin{tabular}{>{$}c<{$}|>{$}c<{$}|>{$}c<{$}|>{$}c<{$}|>{$}c<{$}|>{$}c<{$}|>{$}c<{$}|>{$}c<{$}|>{$}c<{$}}
    \hline
    ~ & \multicolumn{2}{c|}{$\psi_1$} & \multicolumn{2}{c|}{$\psi_2$} &
    \multicolumn{2}{c|}{$\chi_1$} & \multicolumn{2}{c}{$\chi_2$}\\
    \hline
    & \psi_{1L} & \psi_{1R} & \psi_{2L} & \psi_{2R} & \chi_{1L} & \chi_{1R} & \chi_{2L} & \chi_{2R}\\
    \hline
    Q_X & 1 & 1 & 1 & 1& 0& 2& 2 & 0\\
    \hline
    Q_Y & 1 & 2 & -1 & 2 & 2 & 1 &2 & -1\\
    \hline
  \end{tabular}
  \caption{An example of charge assignments for the $\uy\times\ux$ of the 4 Dirac
    fermions. %Charges of the scalar fields $H$ and $\Phi$ are equal to 1. 
    The anomaly coefficient $\kappa$ (Eq.~(\ref{eq:95})) is nonzero and equal to
    6.%  The choice of charges is possible for $m_2 = 0$ in Eq.(\ref{eq:70}).
  }
  \label{tab:choice}
\end{table}

To make the anomalous structure of the Lagrangian~(\ref{eq:70}) more
transparent, we can perform a chiral change of variables, that makes the
fermions vector-like.  Namely, let us start with the term $m_1 \bar\psi_1
e^{i\theta_H\g5}\psi_1$. We want to perform a change of variables to a new
field $\tpsi$, which will turn this term into $m_1 \bar\tpsi_1\tpsi_1$. This
is given by
\begin{equation}
  \label{eq:75}
  \left(\begin{matrix}
    \psi_{1L}\\ 
    \psi_{1R}
  \end{matrix}\right) \to \left(\begin{matrix}
    e^{-\frac i2\theta_H}\tpsi_{1L} \\
    e^{\frac i2 \theta_H}\tpsi_{1R}
  \end{matrix}\right)\quad\text{or} \quad \psi \to e^{-\frac
  i2\g5\theta_H}\tpsi
\end{equation}
so that the Yukawa term becomes $m_1 \bar\tpsi_1\tpsi_1$.  The field
$\tilde\psi_1$ has vector-like charge $x$ with respect to \ux and vector-like
charge $y+\frac12$ with respect to \uy.  As the change of variables is chiral,
it introduces a Jacobian %.  The chiral transformation~(\ref{eq:75}) has a non-trivial Jacobian
 $J_{\psi_1}$~\cite{Fujikawa:1979ay}. The
transformation~(\ref{eq:75}) turns the term $M_1 (\bar \psi_1 e^{i\g5\theta_X}
\chi_1 + c.c.)$ into $M_1 (\bar \tpsi_1 e^{i\g5(\theta_X-\frac{\theta_H}2)}
\chi_1 + c.c.)$. By performing a change of variables from $\chi_1$ to
$\tchi_1$,
\begin{equation}
  \label{eq:78}
  \chi_1 \to e^{-\frac
    i2\g5(\theta_X-\frac{\theta_H}2)}\tchi_1,
\end{equation}
we make the sector $\tpsi_1,\tchi_1$ fully vector-like, and generate two
anomalous Jacobians $J_{\psi_1}$ and $J_{\chi_1}$. Similarly, for the last
term in eq.~(\ref{eq:70}), we perform the change of variables $\chi_2\to
e^{i\theta_X/2}\chi_2$ and $\psi_2\to e^{i\theta_X/2}\psi_2$, generating two
more Jacobians. By computing the Jacobians, one can easily show that
performing the above change of variables for all 4 fermions, we arrive to a
vector-like Lagrangian with the additional term~(\ref{eq:91}).

%%%%%%%%%%%%%%%%%%%%%%%%%%%%%%%%%%%%%%%
\section{Charges in a Realistic $SU(2)\times \uy\times \ux$ Model}
\label{sec:choic-charg-real}

The above example shows us how to construct a realistic model of high-energy
theory, whose low-energy effective action produces the terms~(\ref{eq:48}).
We consider the following fermionic content (iso-index $a=1,2$ marks $SU(2)$
doublets): two left $SU(2)$ doublets $\psi^a_{1L}$ and $\chi^a_{2L}$, two
right $SU(2)$ doublets ${\psi^a_2}_R$ and ${\chi^a_1}_R$, as well as two left $SU(2)$
singlets ${\psi_2}_L$ and ${\chi_1}_L$, and two right $SU(2)$ singlets
${\psi_1}_R$ and ${\chi_2}_R$. The corresponding charge assignments are shown in
Table~\ref{tab:su2}.

\begin{table}
  \centering
  \begin{tabular}{>{$}c<{$}|>{$}c<{$}|>{$}c<{$}|>{$}c<{$}|>{$}c<{$}|>{$}c<{$}|>{$}c<{$}|>{$}c<{$}|>{$}c<{$}}
    \hline
    ~ & \multicolumn{2}{c|}{$\psi_1$} & \multicolumn{2}{c|}{$\psi_2$} &
    \multicolumn{2}{c|}{$\chi_1$} & \multicolumn{2}{c}{$\chi_2$}\\
    \hline
    & \psi^a_{1L} & \psi_{1R} & \psi_{2L} & \psi^a_{2R} & \chi_{1L} & \chi^a_{1R} & \chi^a_{2L} & \chi_{2R}\\
    \hline
    Q_X & x & x & x & x& x-1& x+1& x+1 & x-1\\
    \hline
    Q_Y & y & y+1 & y_1 & y_2 & y+1 & y & y_2 & y_1\\
    \hline
  \end{tabular}
  \caption{Charge assignment for the $SU(2)\times \uy\times\ux$ gauge
    group. Fermions, which are doublets with respect to the $SU(2)$ are marked
    with the superscript ${}^a$. Charges of the SM Higgs field $H$ and of the heavy Higgs $\Phi$ are equal to (1,0) and (0,1) with respect to $\uy\times\ux$.}
  \label{tab:su2}
\end{table}
The Yukawa interaction terms have the form:
\begin{equation}
  \begin{aligned}
    \CL_\text{Yukawa} &= f_1 (\bar\psi^a_{1L} H_a)\psi_{1R} + F_1
    \Bigl(\bar\psi_{1L}^a (\Phi_1 - i\g5\Phi_2)\chi_{1R}^a +
    \mathrm{c.c.}\Bigr)\\
    &+F_2 \Bigl(\bar\psi_{2R}^a
    (\Phi_1 + i\g5\Phi_2)\chi_{2L}^a + \mathrm{c.c.}\Bigr)\\
    & + \tilde F_1 \Bigl(\bar\psi_{1R} (\Phi_1 - i\g5\Phi_2)\chi_{1L} +
    \mathrm{c.c.}\Bigr)+\tilde F_2 \Bigl(\bar\psi_{2L} (\Phi_1 +
    i\g5\Phi_2)\chi_{2R} + \mathrm{c.c.}\Bigr)
\end{aligned}
\label{eq:60}
\end{equation}
where $H$ is the SM Higgs boson and $\Phi_{1,2}$ are $SU(2)\times U(1)_Y$ singlets.
Here again $\langle H\rangle = v \ll \langle \Phi\rangle$, and all states have
heavy masses $\sim F \langle\Phi\rangle$ (plus possible corrections of
order ${\cal O}(fv)$).

The anomaly analysis is similar to the one performed in the previous section.
The only difference being of course two isospin degrees of freedom in the
$SU(2)$ doublets. The resulting choice of charges is shown in
Table~\ref{tab:su2ex} (we do not write the general expression as it is too
cumbersome and provides only an example when $x=-{Q_H}/6$,
$y={Q_\Phi}/2$). %The choice of charges is shown in Table~\ref{tab:su2ex}.
One may check that for this choice of charges the resulting coefficients
$\bc{1,2}$ in the interaction terms~(\ref{eq:48}) are non-zero, which leads to
interesting phenomenology to be discussed in the next section.

\begin{table}
  \centering
  \begin{tabular}{>{$}c<{$}|>{$}c<{$}|>{$}c<{$}|>{$}c<{$}|>{$}c<{$}|>{$}c<{$}|>{$}c<{$}|>{$}c<{$}|>{$}c<{$}}
    \hline
    ~ & \multicolumn{2}{c|}{$\psi_1$} & \multicolumn{2}{c|}{$\psi_2$} &
    \multicolumn{2}{c|}{$\chi_1$} & \multicolumn{2}{c}{$\chi_2$}\\
    \hline
    & \psi^a_{1L} & \psi_{1R} & \psi_{2L} & \psi^a_{2R} & \chi_{1L} & \chi^a_{1R} & \chi^a_{2L} & \chi_{2R}\\
    \hline
    Q_X & -\frac16 & -\frac16 & -\frac16 & -\frac16& -\frac76 & \frac56 & \frac56 & -\frac76\\
    \hline
    Q_Y & \frac12 & \frac32 & -\frac16 & -\frac76 & \frac32 & \frac12 & -\frac76 & -\frac16\\
    \hline
  \end{tabular}
  \caption{Explicit charge assignment for the $SU(2)\times \uy\times\ux$ gauge
    group. 
    %Fermions, which are doublets with respect to the $SU(2)$ are marked
    %with the superscript ${}^a$. Charges of the SM Higgs field $H$ and of the heavy Higgs $\Phi$ are equal to 1.
    }
  \label{tab:su2ex}
\end{table}

\section{Phenomenology}
\label{sec:phenomenology}

The analysis of the previous sections puts us in position to now discuss the phenomenology
of the $X$ boson.  To do this, we  first detail the relevant interactions it has with the SM gauge bosons. 

\begin{figure}[t]
  \centering
  \includegraphics[width=\textwidth]{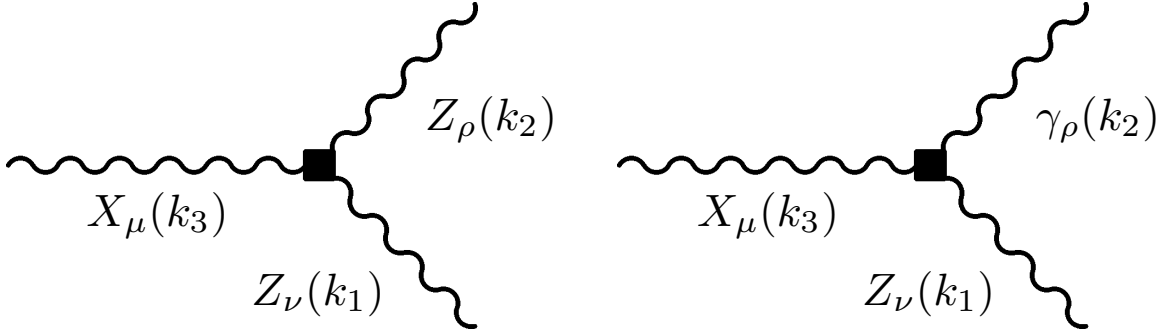} %
  \caption{$\Gamma_{XZZ}$ and $\Gamma_{XZ\gamma}$ interaction vertices, generated by~(\ref{eq:15})}
  \label{fig:vertex}
\end{figure}

The first term in~(\ref{eq:48}) generates two interaction vertices: $XZZ$ and
$XZ\gamma$ (Fig.\ref{fig:vertex}).  In the EW broken phase one can think of
the first term in expression~(\ref{eq:48}) as being simply
\begin{equation}
  \label{eq:15}
  \CL_{XZY} = 
  \bc1 (d\theta_Z + Z) F_Y D\theta_X +
  \mathcal{O}\left(\frac {\partial h}v\right)
\end{equation}
where we parametrized the Higgs doublet as
\begin{equation}
  \label{eq:13}
  H = e^{i (\tau^+ \theta_+(x)+\tau^-\theta_-(x) + (\frac12+\tau^3)\theta_Z)}\left({0 \atop v+h(x)}\right)
\end{equation}
Here the phases $\theta_\pm, \theta_Z$ will be ``eaten'' by $W^\pm$ and $Z$
bosons correspondingly, $v$ is the Higgs VEV and the real scalar field $h$
is the physical Higgs field.

The vertices $\Gamma_{XZZ}$ and $\Gamma_{XZ\gamma}$ are given correspondingly
by%\footnote{Recall that the hypercharge field is related to the $Z$ and photon
%  $\gamma$ via $ Y = -Z\sw + \gamma\cw$.}
\begin{equation}
  \label{eq:40}
  \begin{aligned}
    \Gamma_{XZZ}^{\mu\nu\rho}(k_1,k_2|k_3) &= \frac12\bc1\,\sw
    \epsilon^{\mu\nu\lambda\rho}
    ({k_2}_\lambda -{k_1}_\lambda)  \\
    \Gamma_{XZ\gamma}^{\mu\nu\rho}(k_1,k_2|k_3) &= \hphantom{-} \bc1\,\cw
    \epsilon^{\mu\nu\lambda\rho} {k_2}_\rho
  \end{aligned}
\end{equation}

Similarly to above one can analyze the second term
in~(\ref{eq:48}). It leads to the interaction $XW^+W^-$:
\begin{equation}
% \label{eq:56}
  \Gamma_{XW^+W^-}^{\mu\nu\rho}(k_1,k_2|k_3) = \bc2\,
  \epsilon^{\mu\nu\lambda\rho}
  ({k_2}_\lambda -{k_1}_\lambda)
\end{equation}

The most important relevant fact to phenomenology 
is that the $X$ boson is produced by and
decays into SM gauge bosons.  We shall discuss in turn the production mechanisms and the decay final states of the $X$ boson and then estimate the discovery capability at colliders.

%%%%%%%%%%%%%%%%%%%%%%%%%%%%%%%%%%%%%%%%%%%
\subsection{Production of $X$ boson}

Producing the $X$ boson must proceed via its coupling to pairs of SM gauge
bosons.  One such mechanism is through \emph{vector-boson fusion}, where two
SM gauge bosons are radiated off initial state quark lines and fused into an
$X$ boson: 
\begin{equation}
  pp\to qq'VV'\to qq'X~{\rm or}~VV'\to X~{\rm for~short,}\label{eq:100}
\end{equation}
where $VV'$ can be $W^+W^-$, $ZZ$ or $Z\gamma$. This production mechanism was
studied in ref.~\cite{Kumar:07}.  One of the advantages is that if the decays
of $X$ are not much different than the SM, the high-rapidity quarks that
accompany the event can be used as ``tagging jets" to help separate signal
from the background.  This production mechanism is very similar to what has
been exploited in the Higgs boson literature.

A second class of production channels is through \emph{associated production}:
\begin{equation} 
  pp\to qq'\to V^*\to XV'
\label{eq:1}
\end{equation}
where an off-shell vector boson $V^*$ and the final state $V'$ can be any of
the SM electroweak gauge bosons: $XZ$, $XW^\pm$ or $X\gamma$.  It turns out
that this production class has a larger cross-section than the vector boson
fusion class.  This is opposite to what one finds in SM Higgs phenomenology,
where $VV'\to H$ cross-section is by $\mathcal{O}(10^2)$ greater than $HV'$
associated production. The reason for this is that both vector bosons can be
longitudinal when scattering into $H$, thereby increasing the $VV'\to H$
cross-section over $HV'$. This is not the case for the $X$ boson production,
in which only one longitudinal boson can be present at the vertex. This leads
to a suppression by $\sim (\sqrt{s}/M_V)^2$ of the process~(\ref{eq:100}) as
opposed to the similar process for the Higgs boson.  For LHC energies
($\sqrt{s}\sim 10$~TeV) this suppression is of the order $10^{-4}$.  Without
special longitudinal enhancements, the two body final state $XV'$ dominates
over the three-body final state $qq'X$, which makes the associated
production~(\ref{eq:1}) about 2 orders of magnitude \emph{stronger} than the
corresponding vector-boson fusion. As we shall see below, the decays of the
$X$ boson are sufficiently exotic in nature that background issues do not
change the ordering of the importance of these two classes of diagrams. Thus,
we focus our attention on the associated production $XV'$ to estimate collider
sensitivities.

\begin{figure}[t]
  \begin{tabular}[c]{cc}
    \begin{minipage}[c]{0.5\linewidth}
      \includegraphics[height=1.1\linewidth,angle=-90]{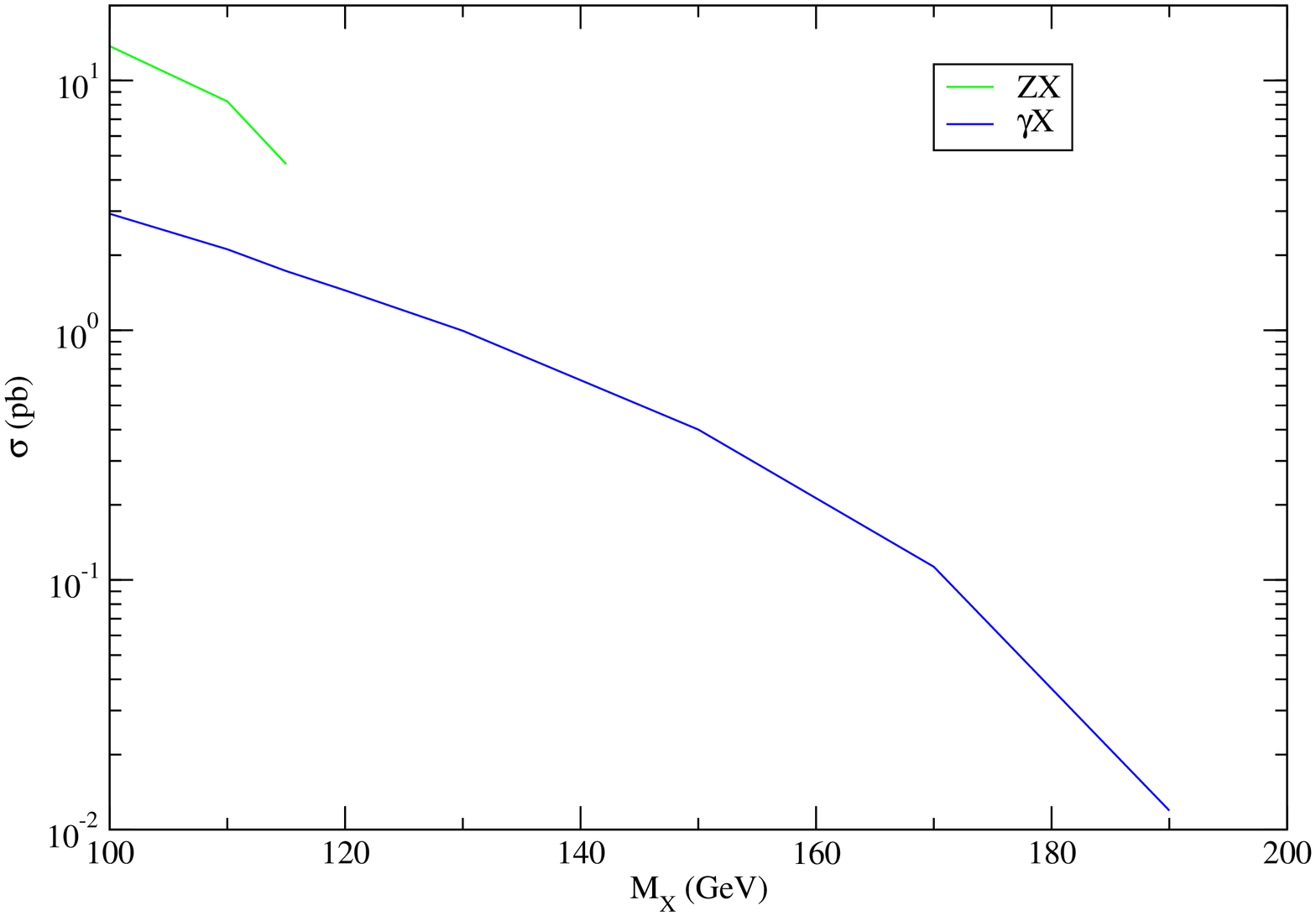}
    \end{minipage}
      &
      \begin{minipage}[c]{0.5\linewidth}
        \includegraphics[height=1.1\linewidth,angle=-90]{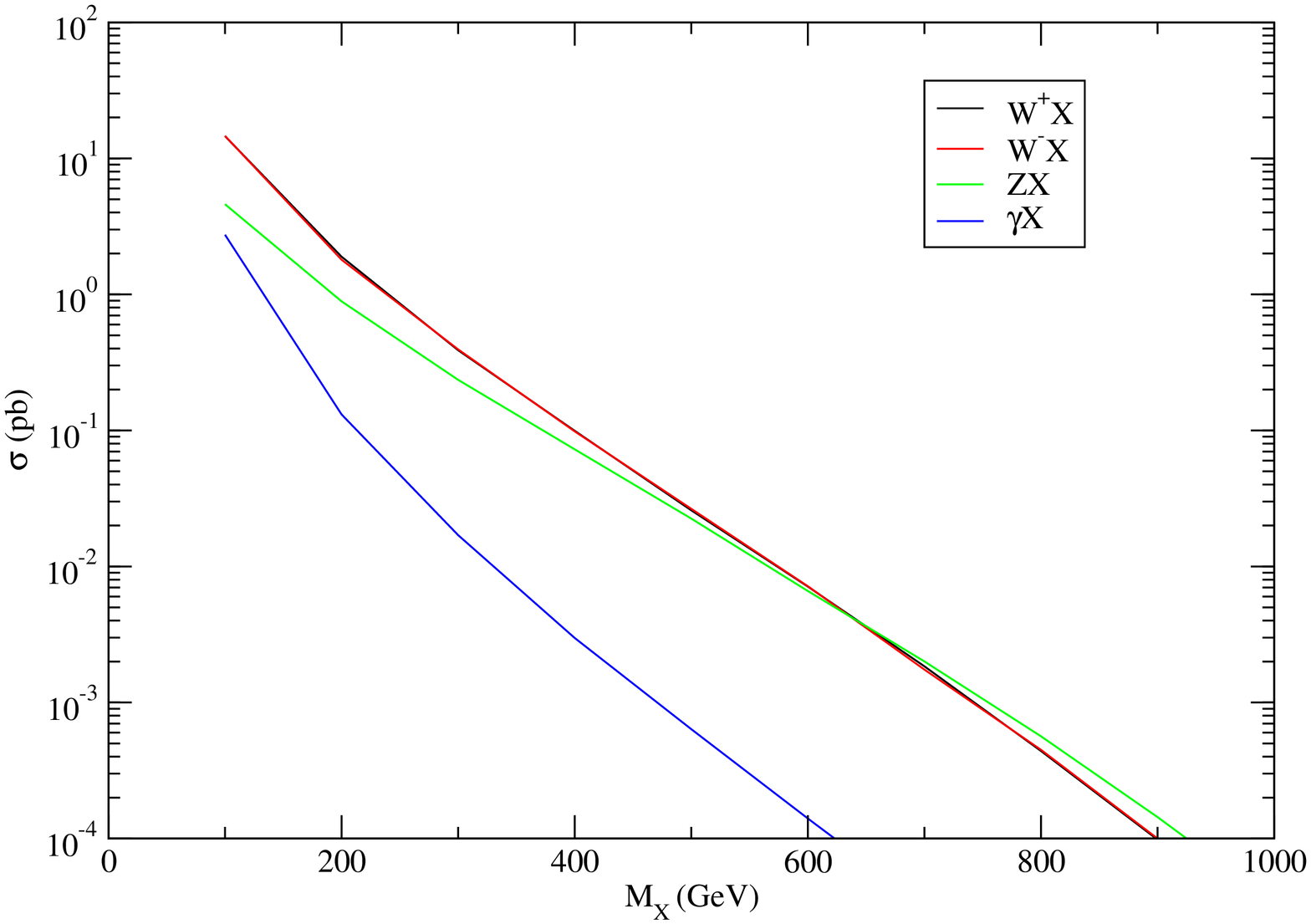}
      \end{minipage}\\
      (a) & (b) \\
    \end{tabular}
    \caption{Production cross-section for $XZ$ and $X\gamma$ at LEP (left) and
      for $XZ$, $X\gamma$, $XW^\pm$ at Tevatron (right panel) vs.\ the $X$
      boson mass. For LEP $\sqrt{s}=200\gev$, and for Tevatron
      $\sqrt{s}=2\tev$. In both cases  $c_1=c_2=1$.}
   \label{fig:lep}
 \end{figure}

\begin{figure}[t]
   \centering
   \includegraphics[width=.4\textheight,angle=-90]{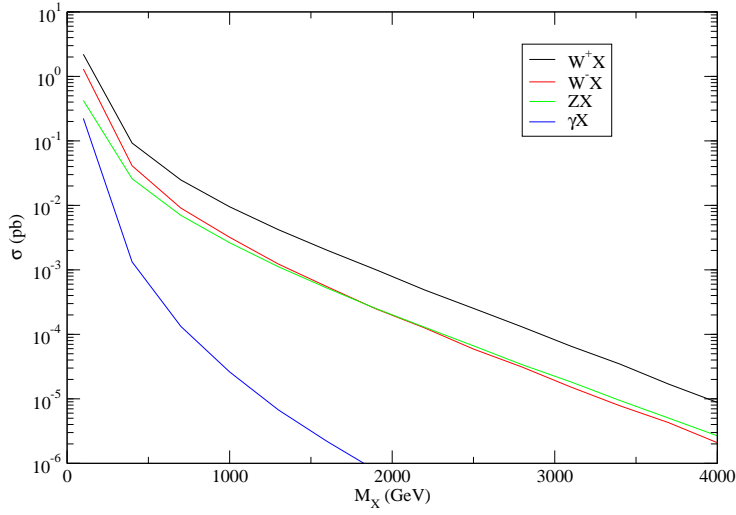}
   \caption{Production cross-section at $\sqrt{s}=14\tev$ LHC of $XV'$ for various $V'=W^\pm, Z, \gamma$ vs.\ the $X$ boson mass with $c_1=c_2=0.1$.}
   \label{fig:lhcsigma}
 \end{figure}

In figs.~\ref{fig:lep} and~\ref{fig:lhcsigma} we plot the production
cross-sections of $XV$ for various $V=W^\pm, Z, \gamma$ at $\sqrt{s}=14\tev$
$pp$ LHC, $\sqrt{s}=2\tev$ $p\bar p$ Tevatron and $\sqrt{s}=200\gev$ $e^+e^-$
LEP.

\begin{figure}[t]
   \centering
   
 \end{figure}

%%%%%%%%%%%%%%%%%%%%%%%%%%%%%%%
\subsection{Decays of $X$ boson}

The $X$ boson decays primarily via its couplings to SM gauge boson pairs.  The  important decay channels are computed from the interaction vertices computed above.
The corresponding decay widths are:
\begin{align}
  \Gamma_{X\to ZZ} & = \frac{\bc1^2 \ssw M_X^3}{192\pi M_Z^2} \left(1-\frac{4
      M_Z^2}{M_X^2}\right)^{5/2} \approx \bc1^2\, (45\gev)
  \left(\frac{M_X}{\mathrm{TeV}}\right)^3 +
  \ldots%\mathcal{O}\left(\frac{M_Z^2}{M_X^2}\right)
  , \nonumber \\
  \Gamma_{X\to W^+W^-} & = \frac{\bc2^2 M_X^3}{48\pi M_W^2}\left(1-\frac{4
      M_W^2}{M_X^2}\right)^{5/2} \label{eq:41} \approx \bc2^2\,
  (1.03\;\mathrm{TeV})\left(\frac{M_X}{\mathrm{TeV}}\right)^3 +\ldots\\
  % \mathcal{O}\left(\frac{M_W^2}{M_X^2}\right)\\
  \Gamma_{X\to Z\gamma} & = \frac{\bc1^2 \ccw M_X^3}{96\pi M_Z^2}
  \left(1-\frac{
      M_Z^2}{M_X^2}\right)^3\left(1+\frac{M_Z^2}{M_X^2}\right)\approx \bc1^2\,
  (307\gev) \left(\frac{M_X}{\mathrm{TeV}}\right)^3 +\ldots,
  % \mathcal{O}\left(\frac{M_Z^2}{M_X^2}\right)
  ,\nonumber
\end{align}
where $\dots$ denote corrections of the order $(M_V/M_X)^2$. The interaction
term of eq.~(\ref{eq:15}) also allows interaction of the $X$ boson with
$\gamma H$ and $ZH$, which are generically small.

 \begin{figure}[t] %www
   \centering
   \includegraphics[width=0.5\textwidth,height=0.6\textheight,angle=-90]{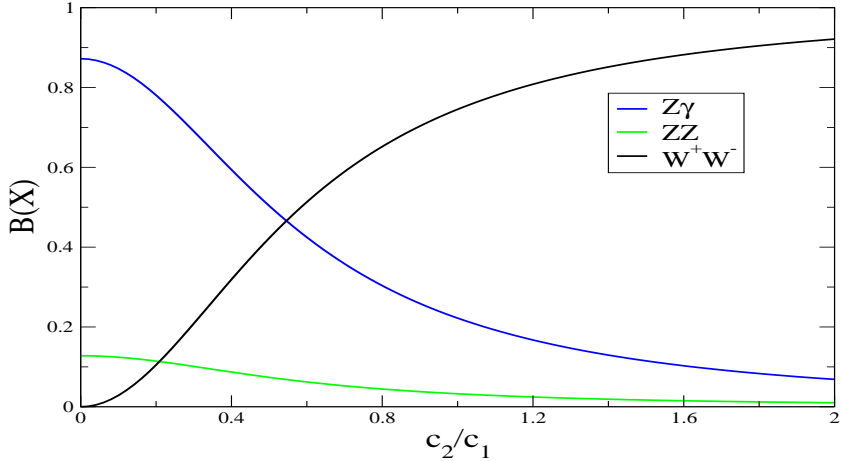} %
   \caption{Branching fractions of $X$ boson decays into $W^+W^-$ (blue), $ZZ$ (yellow-green) and $Z\gamma$ (purple)
as a function of $c_2/c_1$ assuming $M_X\gg M_Z$.}
   \label{fig:XBF}
 \end{figure}

At leading order in $M_Z/M_X$ the decay width into $Z\gamma$ exceeds that of $ZZ$ by
\begin{equation}
  \label{eq:55}
  \frac{\Gamma_{X\to Z\gamma}}{\Gamma_{X\to ZZ}} = 2\frac{\ccw}{\ssw}\approx 6.7
\end{equation}
The branching ratio into $WW$ is the largest over much of parameter space where $c_2\gsim c_1$, and  exceeds that of $ZZ$ by
\bea
  \label{eq:57}
  \frac{\Gamma_{X\to W^+W^-}}{\Gamma_{X\to ZZ}}=
  \frac{4}{\ssw}\,\frac{\bc2^2}{\bc1^2} \approx 17.4\frac{\bc2^2}{\bc1^2}.
\eea
This ratio depends on the a priori unknown ratio of couplings $c_2/c_1$. In Fig.~\ref{fig:XBF} we plot the branching fractions of $X$ into the $WW$ (blue), $Z\gamma$ (purple) and $ZZ$ (yellow-green) as a function of $c_2/c_1$.

Let us compare decay widths~(\ref{eq:41}) with analogous expressions
from~\cite{Kumar:07}. Schematically, decay
widths %~(\ref{eq:41}), (\ref{eq:49})
can be obtained in our case as
\begin{equation}
  \label{eq:50}
  \Gamma_{X\to VV} \sim \bc{1,2}^2 \frac{M_X^3}{M_V^2}
\end{equation}
where we denote by $V$ both $Z$ and $W^\pm$ vector bosons and $M_V =
\{M_Z,M_W\}$.  In case of setup of Ref.~\cite{Kumar:07} the interaction is the
dimension 6 operators, suppressed by the cutoff scale $\Lambda_X^2$. Therefore,
the decay width is suppressed by $\Lambda_X^4$ and the whole expression is
given by
\begin{equation}
  \label{eq:51}
  \Gamma_{X\to VV} \sim \frac{M_X^4}{\Lambda_X^4}\frac{M_X^3}{M_V^2} \frac{M_V^4}{M_X^4} = \frac{M_X^3
    M_V^2}{\Lambda^4}
\end{equation}
The presence of the factor $\frac{M_V^4}{M_X^4}$, appearing in the first
equation of~(\ref{eq:51}), can be explained as follows. The vector boson current
is conserved in the interaction, generated by the higher-dimensional operators
of Ref.~\cite{Kumar:07}. Therefore the corresponding probability for emitting
on-shell $Z$ or $W$ boson is suppressed by the $(\frac{M_V}{E})^4$ where the
energy $E\sim M_X$.  In case of the interaction~(\ref{eq:48}) the vector
current is not conserved in the vertex and therefore such a suppression does
not appear.

%%%%%%%%%%%%%%%%%%%%%%%%%%%%%%%%%%%%%%%%%%%
\subsection{Collider Searches}

Combining the various production modes and branching fractions yields many permutations of final states to consider at high energy colliders.  All permutations, after taking into account $X$ decays, give rise to three vector boson final states such as $ZZZ$, $W^+W^-\gamma$, etc.  The collider phenomenology associated with these kinds of final states is interesting, and we focus on a few aspects of it below.

Our primary interest will be to study how sensitive the LHC is to finding this
kind of $X$ boson.  The limits that one can obtain from LEP 2 and Tevatron are
well below the sensitivity of the LHC, and so we forego a more thorough
analysis of their constraining power. Briefly, in the limit of no background,
the Tevatron cannot do better than the mass scale at which at least a few
events are produced. This implies from fig.~\ref{fig:lep}b that $M_X\gsim
750\gev$ (for $c_i=1$) is inaccessible territory to the Fermilab with up to
$10\, {\rm fb}^{-1}$ of integrated luminosity. The LHC can do significantly
better than this, as we shall see below.

Moving to the LHC, the energy is of course an important increase as is the
planned luminosity.  After discovery is made a comprehensive study programme
to measure all the final states, and determine production cross-sections and
branching ratios would be a major endeavor by the experimental community.
However, the first step is discovery. In this section we demonstrate one of
the cleanest and most unique discovery modes to this theory.  As has been
emphasized earlier and in ref.~\cite{Anastasopoulos:06}, the $X\to \gamma Z$
decay mode is especially important for this kind of theory. Thus, we study
that decay mode. Consulting the production cross-sections results for LHC, we
find that producing the $X$ in association with $W^\pm$ gives the highest
rate. Thus, we focus our attentions on discovering the $X$ boson through
$XW^\pm$ production followed by $X\to \gamma Z$ decay.

The $\gamma Z W^\pm$ signature is an interesting one since it involves all
three electroweak gauge bosons. If the $Z$ decays into leptons, it is
especially easy to find the $X$ boson mass through the invariant mass
reconstruction of $\gamma l^+l^-$.  The additional $W$ is also helpful as it
can be used to further cut out background by requiring an additional lepton if
the $W$ decays leptonically, or by requiring that two jets reconstruct a $W$
mass.

In our analysis, we are very conservative and only consider the leptonic decays of the $Z$ and the $W$. Thus, after assuming $X\to \gamma Z$ decay, $1.4$ percent of $\gamma ZW^\pm$ turn into $\gamma l^+l^- {l'}^\pm$ plus missing $E_T$ events. These events have very little background when cut around their kinematic expectations.  For example, if we assume $M_X=1\tev$ we find negligible background while retaining $0.82$ fraction of all signal events when we making kinematic cuts $\eta(\gamma,l)<2.5$, 
%$\Delta R>0.4$ separation of among all leptons and photons,
$m_{l^+l^-}=m_Z\pm 5\gev$, $p_T(\gamma)>50\gev$, $p_T(l^+,l^-,l')>10\gev$, missing $E_T$ greater than $10\gev$ and $m_{\gamma l^+l^-}>500\gev$.  Thus, for $10\, {\rm fb}^{-1}$ of integrated luminosity at the LHC, when $c_i=1$ ($c_i=0.1$) we get at least five events of this type, $\gamma l^+l^-l'$ plus missing $E_T$, if $M_X>4\tev$ ($M_X>2\tev$).  This would be a clear discovery of physics beyond the SM and would point to a new resonance, the $X$ boson.

\begin{figure}[t]
   \centering
   \includegraphics[width=.6\textwidth,height=.25\textheight]{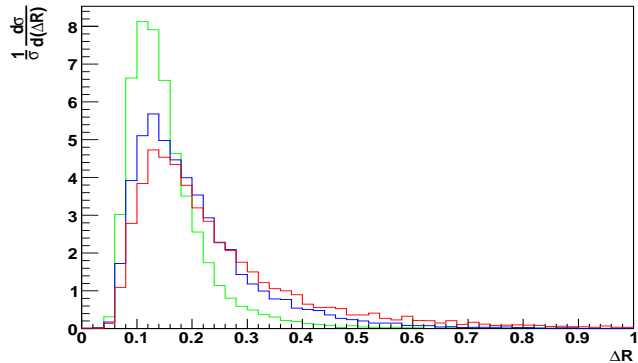} %
   \caption{Distribution of $\Delta R$ of $e^+e^-$ in the $Z$ decays of $WX$ production followed by $X\to 
  \gamma Z$ followed by $Z\to e^+e^-$.  The distributions are for $M_X=500\gev$ (red), $M_X=1\tev$ (blue) and $M_X=2\tev$ (green).}
      \label{fig:3deltar}
 \end{figure}

One subtlety for this signal is the required separation of the leptons from the $Z$ decay in order to distinguish two leptons and be able to reconstruct the invariant mass well. The challenge arises because the $Z$ is highly boosted if its parent particle has mass much greater than $m_Z$, and thus the subsequent leptons from $Z$ decays are highly boosted and collimated in the detector.  One does not expect this to be a problem for $Z\to \mu^+\mu^-$ decays, as muon separation is efficient. Separation of electron and positron in the electromagnetic calorimeter in highly boosted $Z\to e^+e^-$ final states is expected to be more challenging. We do not attempt to give precise numbers of separability for $e^+e^-$. Instead, we only make two relevant comments. First, one is safe restricting to muons. Second, once separability of $e^+e^-$ is better understood, it can be compared with the kinematic distributions of this example to estimate the number of events that are cut out due to the inability to resolve $e^+e^-$. In Fig.~\ref{fig:3deltar} we show the $\Delta R$ separation of $e^+e^-$ for a parent $M_X=500\gev$, $1\tev$ and $2\tev$.  For example, if it turns out that $\Delta R>0.2 \, (0.1)$ is required, then one can expect about $2/3$ ($1/4$) of the $e^+e^-$ events are cut out by this separation criterion.

After discovery, in addition to doing a comprehensive search over all
 possible final states, each individual final state will be studied carefully
 to see what evidence exists for the spin of the $X$ boson. The topology of
 $\gamma Z W^\pm$ exists within the SM for $HW^\pm$ production followed by
 $H\to \gamma Z$ decays. However, the rate at which this happens is very
 suppressed even for the most optimal mass range of the Higgs
 boson~\cite{Djouadi:1996yq}.  A heavy resonance that decays into $\gamma Z$
 would certainly not be a SM Higgs boson, but nevertheless a scalar origin
 would be considered if a signal were found. Careful studying of angular
 correlations among the final state particles can help determine this question
 directly.  For example, distinguishing between the scalar and vector spin possibilities
 of the $X$ boson is possible by carefully analyzing the photon's $\cos\theta_\gamma$ distribution with respect to the $X$ boost direction in $X\to \gamma Z$ decays in the rest frame of the $X$. If $X$ is a scalar its distribution is flat in $\cos\theta$, whereas if it is a vector it has a non-trivial dependence on $\cos\theta$. With enough events (several hundred) this distribution can be filled in, and the spin of the $X$ resonance can be discerned among the possibilities.

\section*{Acknowledgments}
We thank J. Kumar, J. Lykken, F. Maltoni, A. Rajaraman, A. De Roeck for helpful
discussions. I.A. was supported in part by the European Commission under the
ERC  Advanced Grant 226371. O.R. was supported in part by the Swiss National
Science Foundation.

%%%%%%%%%%%%%%%%%%%%%%%%%%%%%%%%%%%
\providecommand{\href}[2]{#2}\begingroup\raggedright\endgroup

%\bibliography{axion,anomalies,numsm,numsm2}
\end{document}